# Alignment Methods and Analysis of Tilt Angle Range for the Data Scanned in Electron Tomography

**Dongwook Kim[a], Kyungtaek Jun[b],***

[a]Department of Mathematics, Atlanta Metropolitan State College, Atlanta, GA, 30310, USA

[b]IM Technology Research Center, 6, Teheran-ro 52-gil, Gangnam-gu, Seoul, 06211, Korea

## Abstract

Electron tomography has been studied in various fields. Various methods have been developed to align projection sets to construct ideally focused reconstruction. In this paper, we present how to align the projection set to distinguish whether it has an ideal sinogram pattern or not by removing translation errors and vertical tilt errors. We also analyze some important properties for certain types of samples to identify whether the reconstruction image can be made through an ideal sinogram pattern. We provide a guideline for how to construct a better reconstruction image by scanning the sample through these properties.

## Introduction

Electron tomography (ET) is becoming an increasingly important tool in various fields such as biological science and materials science for studying the three-dimensional structures [1-10]. ET allows computing three-dimensional reconstructions of objects from their projection recorded at several angles. ET makes it possible to directly visualize the molecular architectures of organelles, cells and complex virus as *in vivo* cellular dynamics [11-13]

The principle of ET is the three-dimensional reconstruction of a specimen from a series of projection images taken with a transmission electron microscope. In ET, an individual biological sample is introduced into the electron microscope and a series of images (the so-called tilt series) is recorded by tilting the sample around a single-axis at different angles, typically over a tilt range of +/- 60 or 70 degrees and at small increments of 1-2 degrees. Typical ET data sets range from 60 to 280 images. Due to the resolution requirements, the image size typically is 2048 x 2048, 4096 x 4096 or even 8192 x 8192 pixels.

Various methods have been developed to align projection sets to create ideally focused reconstruction in tomography [14-19]. A method for making an ideally focused reconstruction using the ideal sinogram pattern has also been developed. In this paper, we present how to align the projection set in ET [20]. It will first show how to distinguish whether it is an ideal sinogram pattern or not by removing translation errors and vertical tilt errors. Second, we analyze some important properties for certain types of samples in the ET. Those are caused by tilt range and it is important to identify whether the reconstruction image can be made through an ideal sinogram pattern. We provide a guideline for how to construct a better reconstruction image by scanning the sample through these properties.

**Analysis for ET sample data**

The electron tomography (ET) data obtained from the planar type samples were obtained by rotating the projection image from side to side based on the widely spread sample. This is similar to selecting an exact projection image when there is a limitation of tilt angle compare to when the data is obtained around a 180 degree rotation angle in tomography. In our ET sample, the projection images become blurred when the tilt angle is large in both negative and positive direction.

Three projection images of a planar type are captured at the tilt angles of -37, 1 and 39 degrees (Fig 1a). This sample includes several distinguishable shapes and computable points that can play a role as fixed points (FPs) including fiducial markers [21-26]. We selected two FPs that are distinguishable in the projection image set. The first fixed point is in the center of the hexagon shape, which is located near the actual rotational axis. This is to reduce the overall image shift when applied to the $T_{0,\varphi}(\theta)$ function for the alignment solution [20]. The second fixed point is in the center of another hexagon that rotates as far as possible from the first fixed point. The height of the rectangle surrounding each fixed point was selected as the height of the figure including the hexagon surrounding each fixed point. This is to make it easy to see how well each part rotates when the tilt angle changes. Figure 1b shows the projection images obtained at the same angle shown in Fig. 1a on the projection image set that is aligned so that the first fixed point is on the $T_{0,\varphi,h_c}(\theta)$ function. We analyzed the trajectory movement of the second fixed point from the projection image set in Fig. 1b. The trajectory of the second fixed point showed the similar pattern with the trajectory of the sample with the vertical tilt error in [20]. Thus, we investigated the trajectory of the FPs at each axial level through the rotation of the projection image, and found that all FPs are on the $T_{r_n,\varphi_n}(\theta)$ in Fig. 1c. Figure 1c shows a rearranged projection image set without translation errors and vertical tilt error, and Fig 1d right panel shows a sinogram from the common layer at 1024 height. The sinogram from the common layer has the ideal sinogram pattern, however, the reconstruction from the data in Fig. 1e cannot be seen at all. Here, we found a new problem in the ET.

**Correlation between tilt angle set range and reconstruction.**

We have made the assumption that there will be a relationship between the tilt angle limit and the object's appearance. To investigate, we made a two-dimensional image sample with a shape of 3000*50 pixels similar to the shape of a real planar type sample. Here, we have checked the correlation with the reconstruction image by excluding the common layer, translation errors, and tilt errors and limiting the tilt angle in the ideal sinogram. Figure 2b left panel shows the ideal sinogram through a tilt angle from -70° to 70° using the radon transform and the sinogram from $0^0$ to $140^0$ is shown in Fig.2b right. In order to see the pattern near tilt angle $0^0$ in Fig.2b right, the maximum value of the density after the angle $70^0$ is limited to the maximum value of the density of the projection image from the angle $70^0$. When the tilt angle change, Δθ, is different, it is necessary to investigate how the reconstruction using a sinogram through a tilt angle from -70° to 70° and the tilt angle through

$0^0$ to $140^0$ differs. For $\Delta\theta = 1^0, 0.5^0$, and $0.15^0$, reconstructions using above different tilt angle ranges mentioned above are shown in Fig. 2c, 2d and 2e, respectively. The top panel is the reconstruction through the tilt angle from -70º to 70º and the bottom panel is the reconstruction through the tilt angle from $0^0$ to $140^0$. In the sinogram, the left is the top part in the reconstruction and the right is the bottom part in the reconstruction. These results illustrate that if the tilt angle range does not include -90º or 90º, increasing the number of projection images will no longer produce a clean image.

Next, we examine what happens in the ET when the change in the shape changes rapidly in the projection image and the density change at a certain angle changes internally. Figure 3a shows an image sample that is round outside but has an ellipse shape inside. We examine the reconstruction image obtained from the portion of the ideal sinogram when the tilt angle passes through the major axis of the ellipse and when it does not. Figure 3b and 3c show sinograms (left) and reconstruction images (right) obtained using limited tilt angle (θ) range, from $-70^0$ to $70^0$ and from $0^0$ to $140^0$ with $\Delta\theta = 1^0$, respectively. Fig. 3d shows a sinogram and its reconstruction through a tilt angle from $0^0$ to $179^0$ with $\Delta\theta = 1^0$. As shown in the yellow circle in the reconstruction image of Fig. 3b and 3c, if the sinogram does not include a tilt angle for a large change in density in the projection, then the error in the construction image becomes larger for tilt angle that is insufficient around a large change in density.

**Discussion**

ET can see ultra-fine material, but there is a limit to the tilt angle limit and the maximum image size to be selected in the projection image set. Thus, the choice of sample for reconstruction images is very important. When creating a reconstruction image, tomography combines the angular parts of each projection image to create the original image. This implies that the projection image should be clear. As shown in the projection image captured at tilt angle $\theta = -67°$ and $77°$ (Fig. 4), the projected image inside the orange colored box is not clear because of the effect that the angle of the planar type sample tilts and the thickness of the sample thickens. It is difficult to obtain a reconstruction image even if it is well aligned with the ideal sinogram pattern. In the previous sample test (Fig. 2), we found that a filtered inverse radon could be accompanied by a larger error in the reconstruction image if there was no projection image area contacting a tilt angle with rapidly changing density in the projection. This is very important since if the tilt angle for a region whose density changes rapidly within the sample is out of tilt angle range, it is impossible to get a good reconstruction image even if there are more projection image by making $\Delta\theta$ samll. In obtaining a tomographic image, the change in density in the projection image is very important regardless of the sample type. For this purpose, it has been proposed that density should be expressed as mass attenuation coefficient (MAC) accumulation. It has also been studied that density is difficult to accurately preserve linearity, and small changes in density can affect reconstructions. This is important for obtaining a clear reconstruction image, however it is also a requirement to use the concept of center of attenuation.

In this paper, we have also found that reconstructions cannot be done if density changes occur in a particular projection image and the projection image of the tilt angle is missed. This implies that reconstruction is difficult if a clean projection image cannot be obtained at an angle of $90^0$ for the planar type sample. In order to obtain a planar type sample in ET, it is necessary to select the projection image to pass through the side of the sample by cutting it into a part that can be cleaned out. We used a test sample to test how the reconstruction image differs by adding a projection image for the angle of projection from $110^0$ to $250^0$, which is the opposite range of $-70^0$ to $70^0$. However, the results are the same as what we used the previous tilt angle range. It has been studied that the reconstruction using a tilt angle range from $0^0$ to $360^0$ is better than using the tilt angle range from $0^0$ to $180^0$ [27-33]. This means that if the projection set does not have an ideal sinogram pattern, then the average error is reduced by increasing the number of projection images and there is no difference in ideally focused reconstruction for samples with ideal sinogram patterns. It is important to obtain a clear projection image to obtain a better reconstruction, but it is difficult to obtain a correct reconstruction if there is no projection image for a tilt angle with a large change in density. This can be applied not only to ET but also to actual CT scan. By selecting a projection image around the tilt angle of a projection image with a large density change in CT, a clearer image can be obtained from a projection image set having a fixed number of images.

It is important to find a common layer in order to obtain a better quality reconstruction image. It has been introduced how to remove translation errors and vertical tilt error and how to calculate the angle of parallel tilt error [20]. To identify if there are tilt errors in a sample, it can be distinguished by placing one fixed point on the projected virtual rotational axis using the virtual focusing method [20] and placing the other fixed point on the other trajectory. In the planar type sample, the second fixed could be applied to $T_{r_n,\varphi_n,h_n}(\theta)$, making it an ideal sinogram pattern. If the projection set has a parallel tilt error, there is no common layer. In this case, only an overall optimal solution is possible. According to the method presented in this paper, the optimal solution is near the virtual rotational axis. To get a clearer view of the other parts in the sample, it is possible to apply a similar approach to the point of the part that we want to see to $T_{r_n,\varphi_n,h_n}(\theta)$ function.

## Method

### Virtual Focusing Method using Fixed Point

The circular trajectory of a point p in the real space corresponds to a curve drawn by the sinusoidal function in the sinogram. The function is given by [20]

$$T_{r,\varphi}(\theta) = r * \cos(\theta - \varphi), 0 \leq \theta < 180° \quad (1)$$

Where $r$ is the distance between the rotation axis and the point $p$. $\theta$ is the projecting angle, and $\varphi$ is the angle between the line $\overleftrightarrow{Op}$ and the orthogonal line to the tilt angle at $\theta = 0$.

In ideal cases, the center $O$ is converted to $T_{0,\varphi}$ in the sinogram, but not in the actual sinogram. $T_{r,\varphi}$ is a function that shows how a specific point $p$ in the real space moves on the sinogram. In other words, if a point $p$ in the solid specimen rotates on the stage and the projected curve drawn by the movement of $p$ for each angle is the same as the sinusoidal curve by $T_{r,\varphi}$ in the sinogram, then the projected trajectories of other points in the specimen should satisfy the projected curves by $T_{r_n,\varphi_n}$.

Let us first consider the points in the sample that can play a role as fixed points. Consider a point near the vertical line of the CCD for easy calculation. Let $T_{0,\varphi,h}$ (the projection line of the virtual rotation axis) be the center of the CCD and place the first fixed point at the appropriate $h_c$ so that the image is not cut off (see Fig. 1 and Fig. 5). In a projection image set without tilt errors, the trajectory of the fixed point appears as a straight line in the CCD. Also, when considering the union of common layers surrounding a specific part of a sample, a specific part at a certain height is drawn in a rotational orbit. Through this process, we found that the projection image has a tilt error since the trajectory of the second fixed point changes in height in the sample. To correct the tilt errors, rotate the second fixed point to a certain height in the projection image. However, this can only correct vertical tilt error. Now, we should look at the trajectory of the second fixed point through the rotated projection image set. If the second fixed point satisfied the function of $T_{r,\varphi}(\theta)$ in its common layer, all remaining points will satisfy the sinusoidal curve in the sinogram, which is the ideal sinogram pattern. If the second fixed point does not satisfy $T_{r,\varphi}(\theta)$, it is due to the parallel tilt error, and the trajectory of this trajectory of the fixed point will show the shape of ellipse on the CCD. In this case, the overall optimal reconstructions are possible. To see a specific region more clearly, it is possible to place the region on $T_{0,\varphi}(\theta)$,

**References**

1. Lucic V, Forster F, Baumeister W. Structural studies by electron tomography: from cells to molecules. *Annual Review of Biochemistry*. **2005**; 74:833-865

2. Fernández JJ, Sorzano COS, Marabini R, Carazo JM. Image processing and 3D reconstruction in electron microscopy, *IEEE Signal Processing Magazine*. **2006**; 23(3):84–94

3. Frank J, ed. Electron Tomography: Methods for Three-Dimensional Visualization of Structures in the Cell. *Springer*; **2006**

4. Weyland, M., Midgley, P. A. & Thomas, J. M. Electron Tomography of Nanoparticle Catalysts on Porous Supports: A New Technique Based on Rutherford Scattering. *J. Phys. Chem*. B **105**, 7882–7886 (2001).

5. Inoue, T. et al. Electron tomography of embedded semiconductor quantum dot. Applied Physics Letters 92, 031902 (2008).

6. Li, S. et al. Interplay of Three-Dimensional Morphologies and Photocarrier Dynamics of Polymer/TiO2 Bulk Heterojunction Solar Cells. *J. Am. Chem. Soc* **133**, 11614–11620 (2011).

7. Carriazo, D. et al. Formation Mechanism of LiFePO4 Sticks Grown by a Microwave-Assisted Liquid-Phase Process. *Small* **8**, 2231–2238 (2012).

8. Keller, L. M. et al. Characterization of multi-scale microstructural features in Opalinus Clay. *Microporous and Mesoporous Materials* **170**, 83–94 (2013).

9. Scott, M. C. et al. Electron tomography at 2.4-angstrom resolution. *Nature* **483**, 444–447 (2012)

10. Xu, R. et al. Three-dimensional coordinates of individual atoms in materials revealed by electron tomography. *Nature Materials* **14**, 1099–1103 (2015).

11. Medalia O, Weber I, Frangakis AS, Nicastro D, Gerisch G, Baumeister W. Macromolecular architecture in eukaryotic cells visualized by cryoelectron tomography. *Science*. **2002**; 298:1209-1213.

12. Beck M, Förster F, Ecke M, Plitzko JM, Melchior F, Gerisch G, Baumeister W, Medalia O. Nuclear pore complex structure and dynamics revealed by cryoelectron tomography. *Science*. **2004**;306:1387-1390

13. Brandt F, Etchells SA, Ortiz JO, Elcock AH, Hartl FU, Baumeister W. The native 3D organization of bacterial polysomes. *Cell*. **2009**; 136:261-271

14. Pan, Y., De Carlo, F. & Xiao, X. Automatic detection of rotational centers using GPU from projection data for micro-tomography in synchrotron radiation. *Proc. SPIE,* **8313***,* 831328 (2012).

15. Gürsoy, D., De Carlo, F., Xiao, X. & Jacobsen, C. TomoPy: a framework for the analysis of synchrotron tomographic data. *J. Synchrotron Rad.* **21**, 1188–1193 (2014).

16. Donath, T., Beckmann, F. & Schreyer, A. Automated determination of the center of rotation in tomography data. *J. Opt. Soc. Am. A,* **23**, 1048–1057 (2006).

17. Parkinson, Y. D., Knoechel, C., Yang, C., Larabell, C. A., & Le Gros, M.A. Automatic alignment and reconstruction of images for soft X-ray tomography. *J. Struct. Biol.***177**, 259–266 (2012).

18. Yang, Y. et. Al. Registration of the rotation axis in X-ray tomography. *J. Synchrotron Rad*., **22**, 452-457 (2015)

19. Vo, N. T., Drakopoulos, M., Atwood, R. C. & Reinhard, C. Reliable method for calculating the center of rotation in parallel-beam tomography. *Opt. Express,* **22**, 19078 (2014).

20. Jun, K. and Yoon, S. Alignment Solution for CT Image Reconstruction using Fixed Point and Virtual Rotation Axis. *Sci. Rep.* 7, **41218**; doi: 10.1038/srep41218 (2017).

21. Kremer, J.R., Mastronarde, D.N. & McIntosh, J.R. Computer visualization of three-dimensional image data using IMOD.*J. Struct. Biol.* **116**, 71–76 (1996).

22. Sorzano, C.O.S. et al. XMIPP: a new generation of an open-source image processing package for electron microscopy. *J. Struct. Biol.* **148**, 194–204 (2004).

23. Amat, F. et al. Markov random field based automatic image alignment for electrontomography. *J. Struct. Biol.* **161**, 260–275 (2008).

24. Chen, H., Hughes, D.D., Chan, T.A., Sedat, J.W. & Agard, D.A., IVE (Image Visualization Environment): a software platform for all three-dimensional microscopy applications. *J. Struct. Biol.* **116**, 56–60 (1996).

25. Liu, Y., Penczek, P.A., Mcewen, B.F. & Frank, J.A marker-free alignment method for electron tomography. *Ultramicroscopy* **58**, 393–402 (1995).

26. Zheng, S.Q. et al. UCSF tomography: an integrated software suite for real-time electron microscopic tomographic data collection, alignment, and reconstruction. *J. Struct. Biol.* **157**, 138–147

(2007).

27. Carrascosa, J.L., Chichon, F.J., Pereiro, E., Rodriguez, M.J., Fernandez, J.J., Esteban, M., Heim, S., Guttmann, P., Schneider, G., 2009. Cryo-X-ray tomography of vaccinia virus membranes and inner compartments. *J. Struct. Biol*. **168**, 234–239.

28. Le Gros, M.A., McDermott, G., Larabell, C.A., 2005. X-ray tomography of whole cells. *Curr. Opin. Struct. Biol*. **15**, 593–600.

29. McDermott, G., Le Gros, M.A., Knoechel, C.G., Uchida, M., Larabell, C.A., 2009. Soft Xray tomography and cryogenic light microscopy: the cool combination in cellular imaging. *Trends Cell Biol*. **19**, 587–595.

30. Meyer-Ilse, W., Hamamoto, D., Nair, A., Lelievre, S.A., Denbeaux, G., Johnson, L., Pearson, A.L., Yager, D., Legros, M.A., Larabell, C.A., 2001. High resolution protein localization using soft X-ray microscopy. *J Microsc-Oxford* **201**, 395–403.

31. Schneider, G., Guttmann, P., Heim, S., Rehbein, S., Mueller, F., Nagashima, K., Heymann, J.B., Muller, W.G., McNally, J.G., 2010. Three-dimensional cellular ultrastructure resolved by X-ray microscopy. *Nat. Methods* **7**, 985-U116.

32. Uchida, M., McDermott, G., Wetzler, M., Le Gros, M.A., Myllys, M., Knoechel, C., Barron, A.E., Larabell, C.A., 2009. Soft X-ray tomography of phenotypic switching and the cellular response to antifungal peptoids in Candida albicans. *Proc. Natl Acad. Sci. USA* **106**, 19375–19380.

33. Uchida, M., Sun, Y., McDermott, G., Knoechel, C., Le Gros, M.A., Parkinson, D., Drubin, D.G., Larabell, C.A., 2011. Quantitative analysis of yeast internal architecture using soft X-ray tomography. *Yeast* **28**, 227–236.

**Figures**

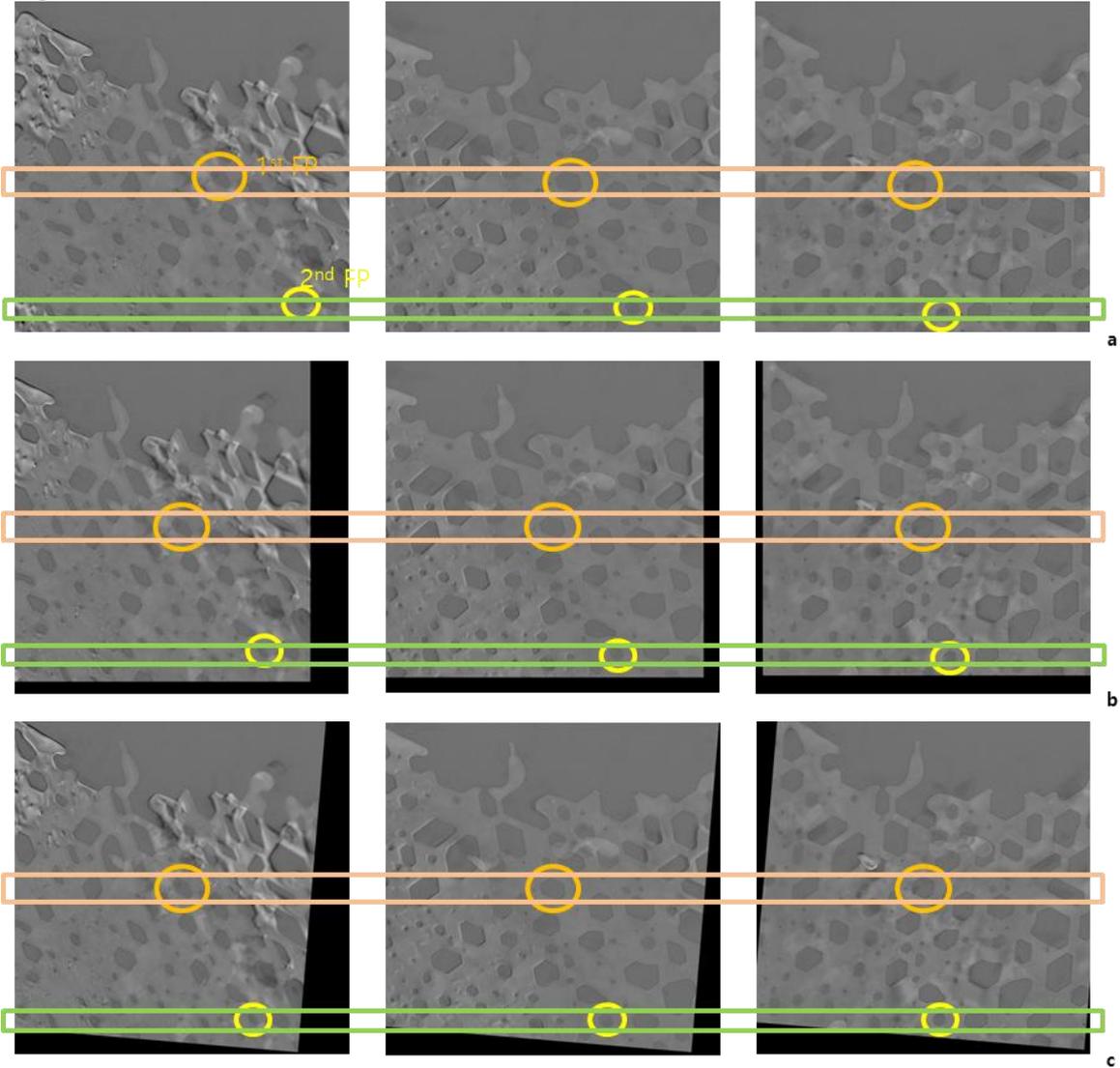

a

b

c

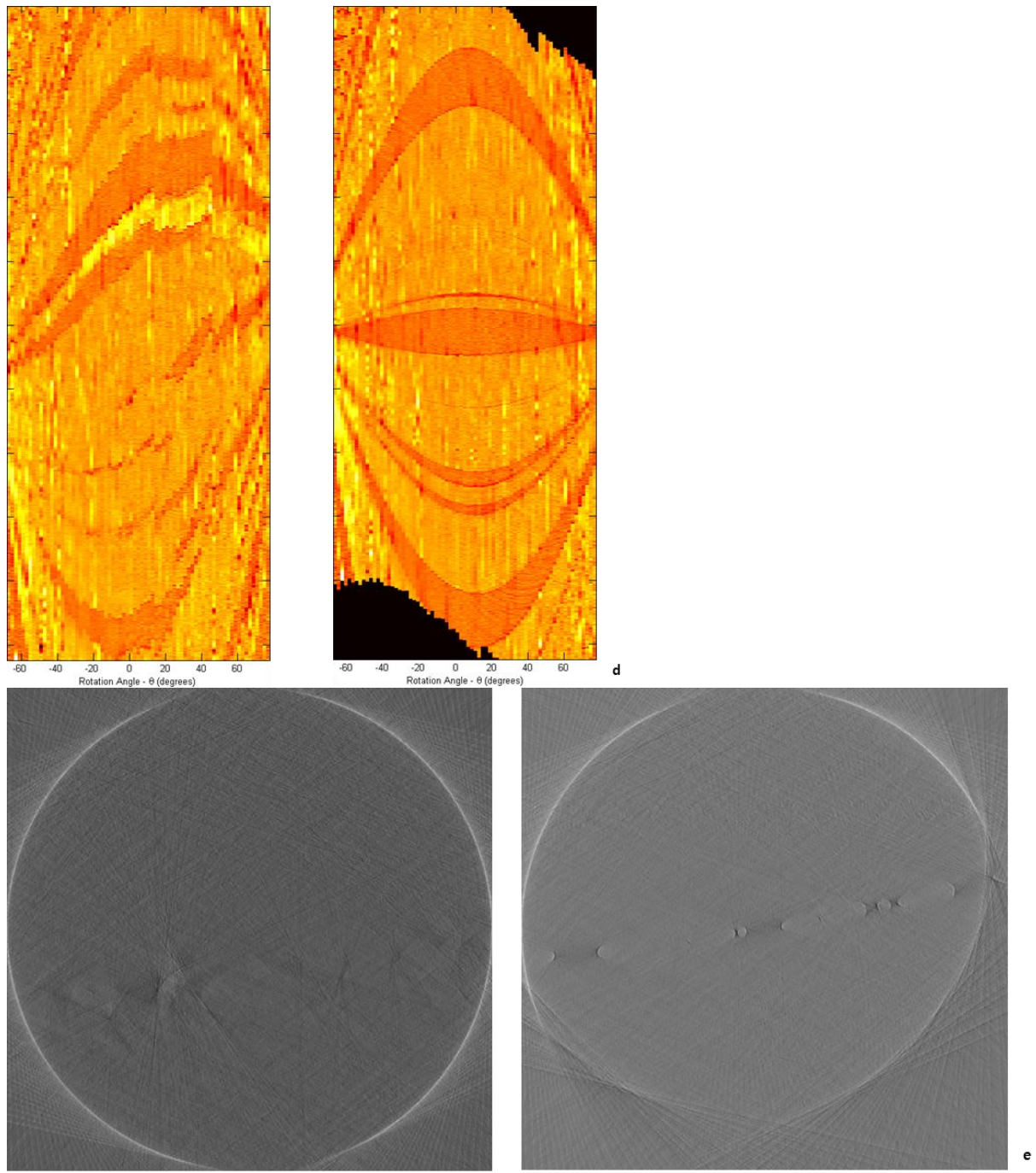

Figure 1. Projection image set of Planar-type samples in ET. The data for the sample was obtained from NSLS II and captured at tilt angle of $-67^0$ and $-77^0$ with a small increment of 2 degrees. The size of each projection image is 2048 x 2048 pixels.

a. Three projection images captured at tilt angle of $-37°$, $1°$ and $39°$.
b. Projection images with no translation errors by making the first fixed point equal height on the CCD and placing it on the projected virtual rotational axis.
c. Projection images were rotated based on the first fixed point to fit the second fixed point to the same

axial level and the vertical tilt error was removed from the projection image set. When we take one common layer according to the height, all distinguishable parts are on the common layer and move along $T_{r,\varphi}(\theta)$. This implies that there is no parallel tilt error and the modified projection set is an ideally arranged projection set.

d. The original sinogram at 1024 axial level (left). The sinogram with ideal pattern at common level of 2024 height in the ideally arranged projection set (right).

e. none ideally focused reconstruction (left) and ideally focused reconstruction (right). Although reconstructed using an ideal sinogram pattern, only the shape of the void space can be distinguished. The projection set of Planar-type in ET is expected to have specific problems.

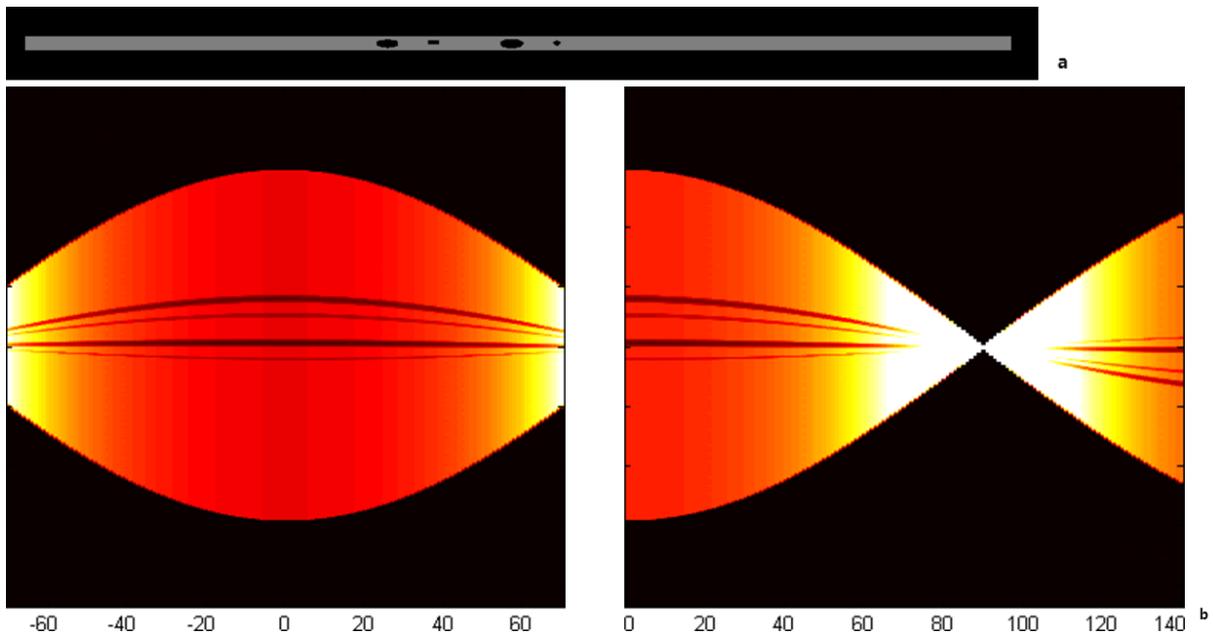

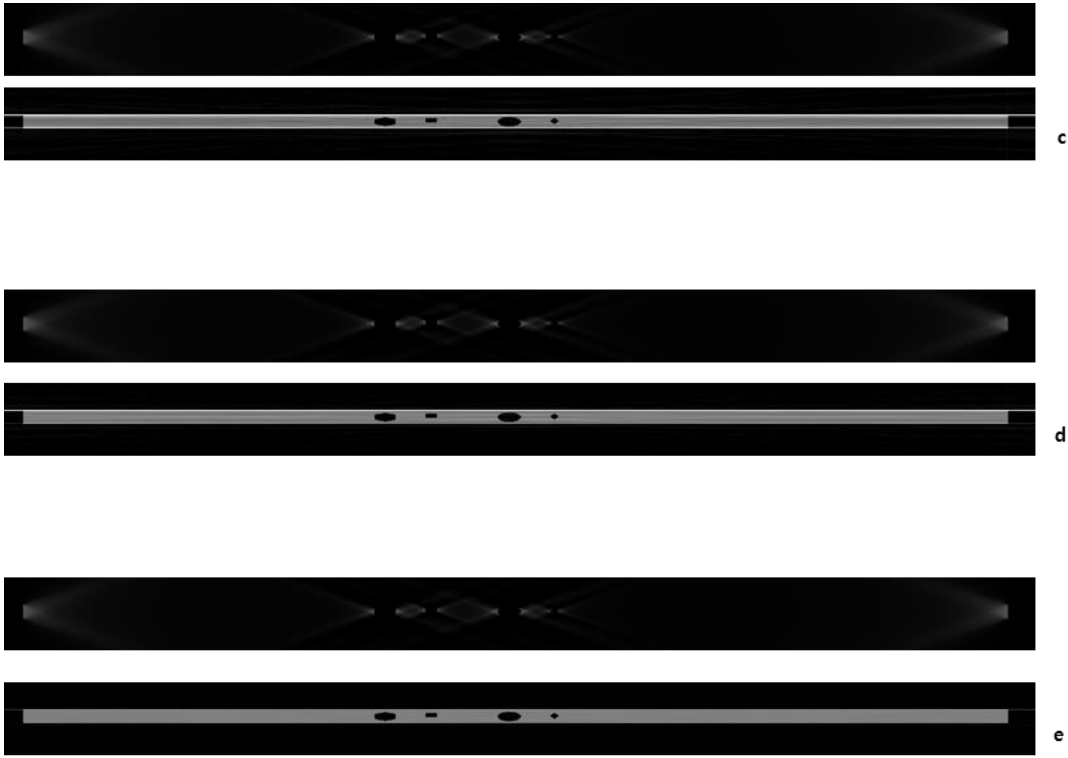

Figure 2. Reconstructions based on the limit of the tilt angle and the external shape of the sample in an image sample. In the projection image, the density changes greatly at the angle $90^0$. If the tilt angle range does not include $90^0$, no matter how much the number of projection images can be increased, it is not possible to obtain a clear reconstruction.

a. Image sample 3000*50

b. Sinograms according the tilt angle. A sinogram with the tilt angle range is from $-70^0$ to $70^0$ (left) and a sinogram with the angle range is from $0^0$ to $140^0$ (right)

c. Reconstructions using different tilt angle ranges with $\Delta\theta = 1°$. A reconstruction through tilt angle ranges from $70^0$ to $70^0$ (top) and a reconstruction through tilt angle ranges from $0^0$ to $140^0$ (bottom)

d. Reconstructions using different tilt angle ranges with $\Delta\theta = 0.5°$. A reconstruction through tilt angle ranges from $70^0$ to $70^0$ (top) and a reconstruction through tilt angle ranges from $0^0$ to $140^0$ (bottom)

e. Reconstructions using different tilt angle ranges with $\Delta\theta = 0.15°$. A reconstruction through tilt angle ranges from $70^0$ to $70^0$ (top) and a reconstruction through tilt angle ranges from $0^0$ to $140^0$ (bottom).

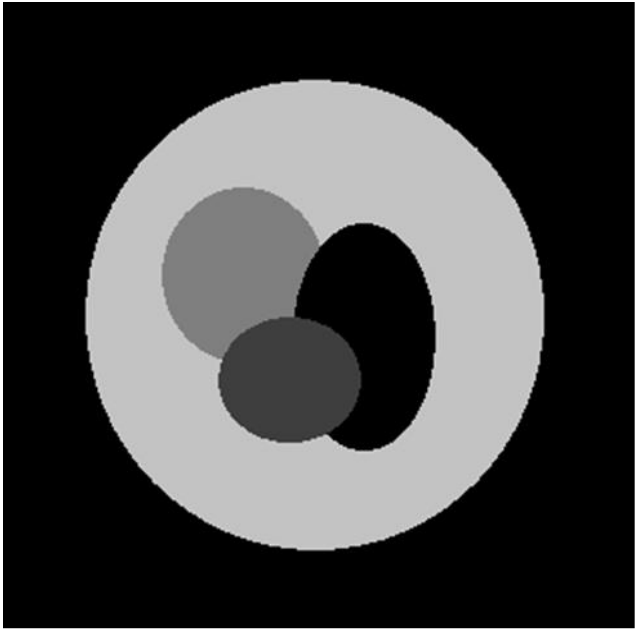

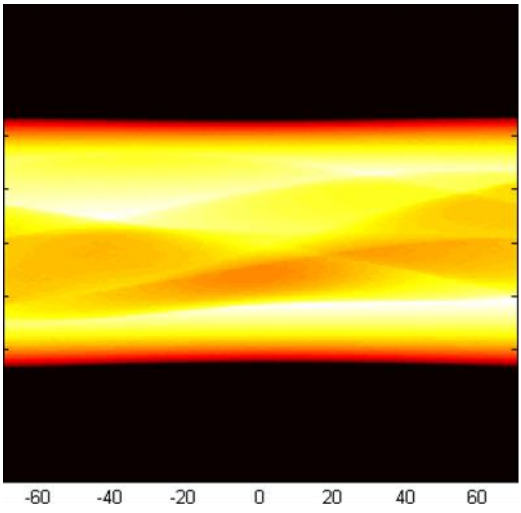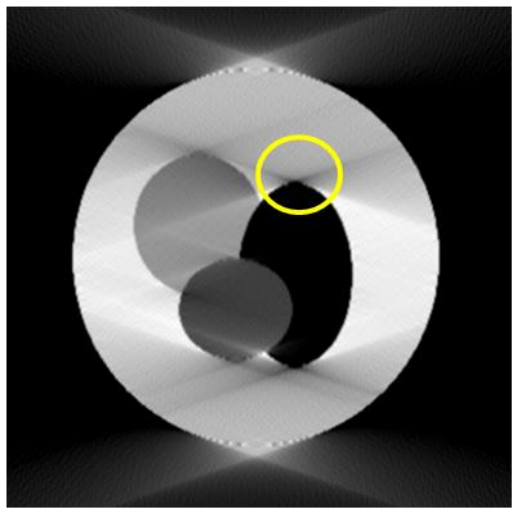

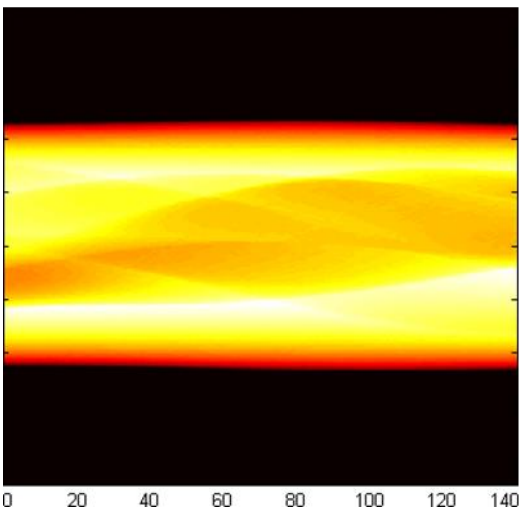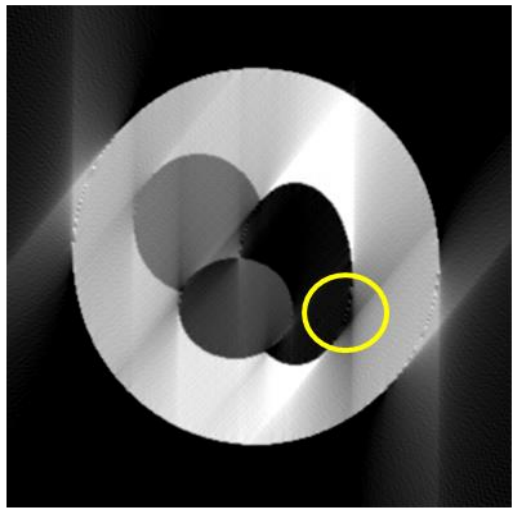

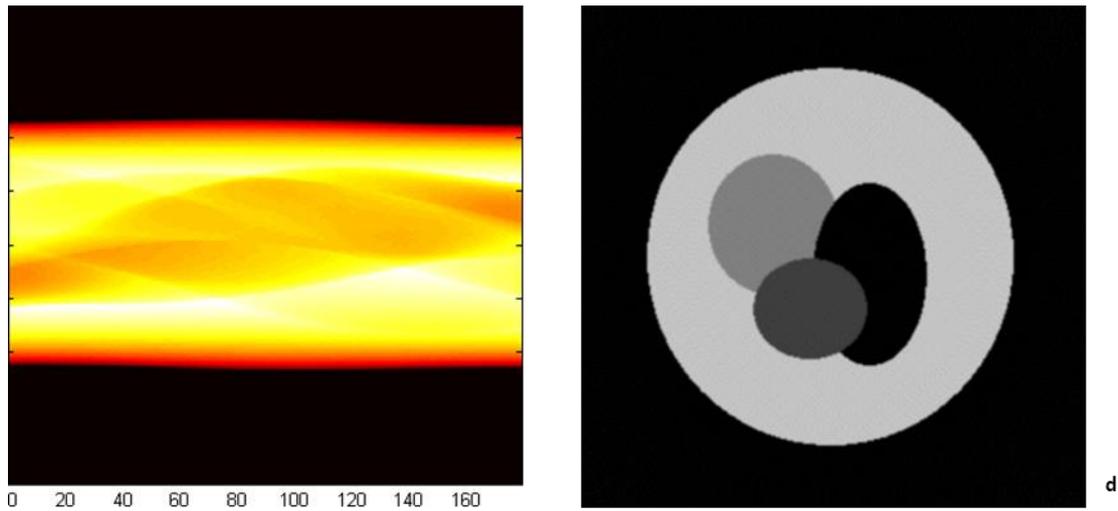

Figure 3. A reconstruction image sample according to the limitation of tilt angle range and internal shape of the sample.

a. A sinogram and reconstruction image with tilt angle range from $-70°$ to $70°$ with $\Delta\theta = 1°$.
b. A sinogram and reconstruction image with tilt angle range from $0°$ to $140°$ with $\Delta\theta = 1°$.
c. A sinogram and reconstruction image with tilt angle range from $0°$ to $179°$ with $\Delta\theta = 1°$.

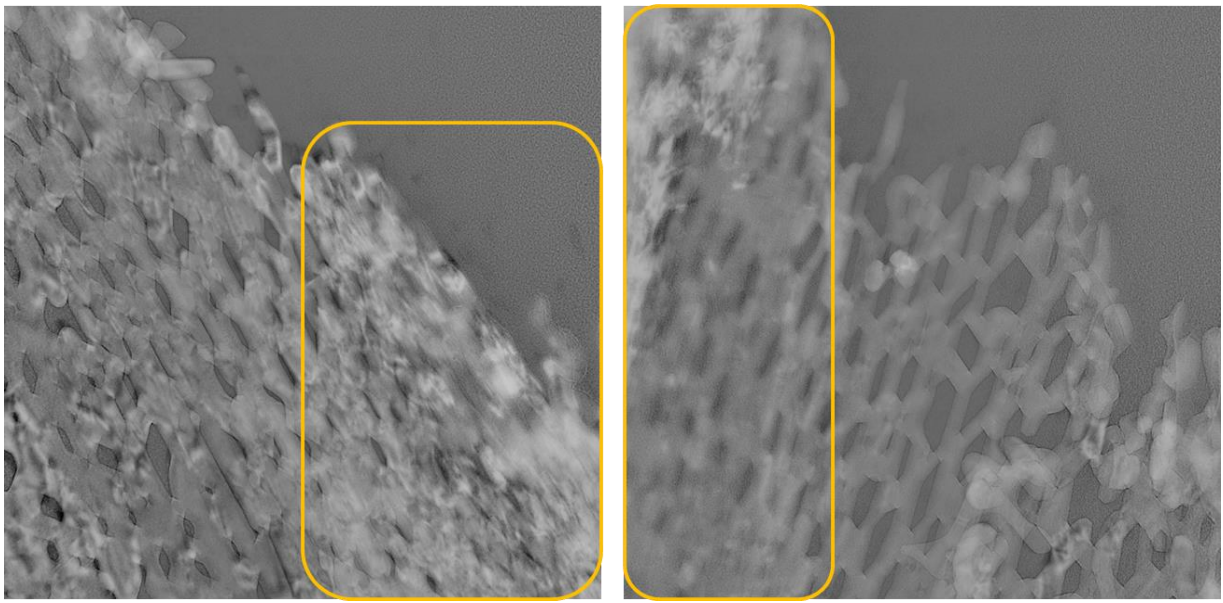

Figure 4. Projection images of a planar-type sample with different tilt angle. The inside of the orange box becomes blurred due to the change in tilt angle. The image capture at tilt angle $-67$ (left) and at $77°$ (right)

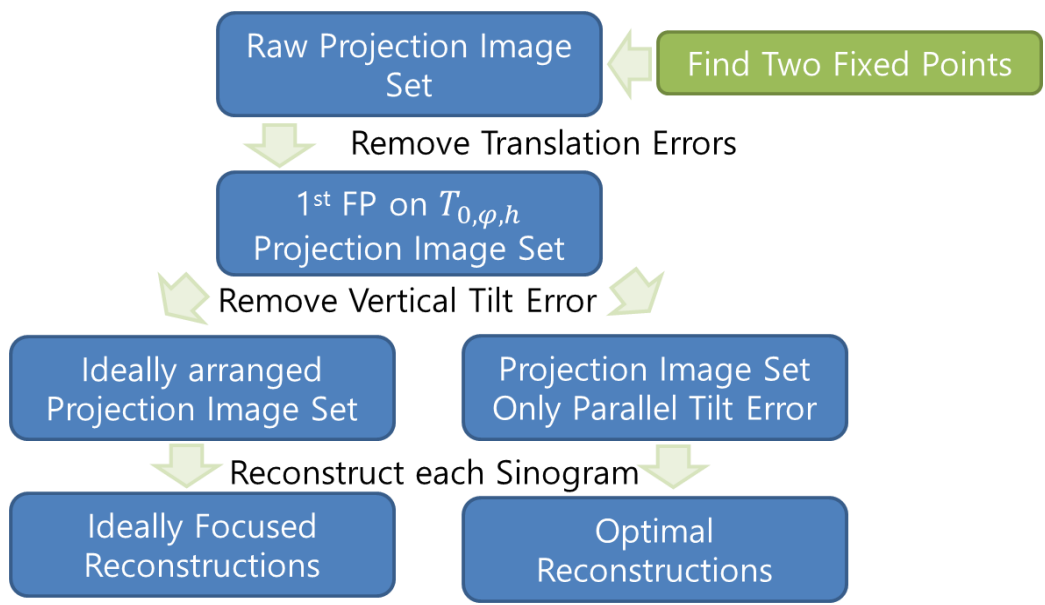

Figure 5. An overview of the methodology we use for alignment of ET projection image set. First, find two fixed points and modify the translation error through the first fixed point. And then, by placing the second fixed point at a specific layer, an image set with only vertical tilt errors can be made into an ideal sinogram pattern. If the sinogram after modifying the vertical tilt error does not have an ideal pattern, this is due to the parallel tilt error, and an overall optimal reconstruction is possible by finding the elliptical trajectory of the second fixed point on the CCD.